\documentclass[prl,twocolumn,showpacs,preprintnumbers,amsmath,amssymb]{revtex4}
\newcommand{\ket}[1]{\left | #1 \right\rangle}

\usepackage[dvips]{graphics}
\usepackage{tikz}

\begin{document}

\title{The Aharonov-Bohm phase is locally generated \\(like all other quantum phases)}

\author{Chiara Marletto and Vlatko Vedral}
\affiliation{Clarendon Laboratory, University of Oxford, Parks Road, Oxford OX1 3PU, United Kingdom and\\Centre for Quantum Technologies, National University of Singapore, 3 Science Drive 2, Singapore 117543 \\
Department of Physics, National University of Singapore, 2 Science Drive 3, Singapore 117542\\
ISI Foundation, Via Chisola, 5, 10126, Turin, Italy}

\date{\today}

\begin{abstract} 
\noindent In the Aharonov-Bohm (AB) effect, a superposed charge acquires a detectable phase by enclosing an infinite solenoid, in a region where the solenoid's electric and magnetic fields are zero. 
Its generation seems therefore explainable only by the local action of gauge-dependent potentials, not of gauge-independent fields. This was recently challenged by Vaidman, who explained the phase by the solenoid's current interacting with the electron's field (at the solenoid). Still, his model has a residual non-locality: it does not explain how the phase, generated at the solenoid, is detectable {\sl on the charge}. In this paper we solve this non-locality explicitly, by quantising the field. We show that the AB phase is mediated locally by the entanglement between the charge and the photons, like all electromagnetic phases. We also predict a gauge-invariant value for the phase difference at each point along the charge's path. We propose a realistic experiment to measure this phase difference locally, by partial quantum state tomography on the charge, without closing the interference loop. 
\end{abstract}


\maketitle                           

In the Aharonov-Bohm (AB) effect, a charge $q$ is superposed across two paths enclosing a magnetic field ${\bf B_0}$, usually produced by an infinite solenoid, \cite{AHBO1} (figure 1). The AB phase is the phase difference between the paths, $\Delta\phi_{AB}=\frac{q}{\hbar}\oint_{S}{\bf B_0}\cdot {\bf ds}$, $S$ being the surface enclosed by the paths. 

The AB phase has been considered anomalous for various reasons, \cite{WAL, VAI, NOR, KAN, LAT1, LAT2}. In the semiclassical model where a classical background field interacts with a quantum charge, the phase appears to be non-locally generated. The classical electromagnetic (EM) field is zero where the wave-function of the charge is non-zero, so the phase cannot occur by the EM field acting locally on the charge. One can still explain it via local action on the charge, but only via the vector potential, ${\bf A}$ (${\bf B_0}={\bf \nabla}\wedge {\bf A}$), whereby $\Delta\phi_{AB}=\frac{q}{\hbar}\oint{\bf A}\cdot {\bf dl}$. This is also problematic, as the vector potential is not a physical observable.  Hence the AB phase has been considered different from all other quantum EM phases. 

Here we expose an additional, crucial consideration, proving that the AB phase no more problematic than any EM quantum phase. We notice two separate problems (so far confused): i) whether the AB phase is generated via local interaction of gauge-independent fields with charges; ii) whether the AB phase is locally acquired by the charge along its path. By `locally acquired' we mean that for any two points, ${\bf r_L}$ on one branch of the superposition, ${\bf r_R}$ on the other, there is a gauge-independent phase difference, detectable by measuring observables of the charge locally at each point, via tomography.  
Our paper addresses problem (ii), proposing an experimental scheme to test our predictions. A solution to problem (i) was proposed by Vaidman, \cite{VAI}. In his model, the AB phase is generated by the EM field  {\sl  of the charge} acting on the solenoid's charges. Vaidman's key idea is that the phase is generated by the solenoid being reversibly entangled with the charge. Kang proposed a Lagrangian bearing out Vaidman's model, \cite{KAN, KIM} (see also \cite{SAGO}, \cite{PAB}). We call these models {\sl field-based} (the fields couple directly with charges) as opposed to the {\sl potential-based} model, where the charge is coupled to the potential $A$. Pearle and Rizzi \cite{PERI} provided a unified quantum treatment, explaining for each model which of the three elements (solenoid, charge and EM potential) is treated classically or quantum-mechanically. 

\begin{figure}[htb]
\centering
\includegraphics[width=85mm]{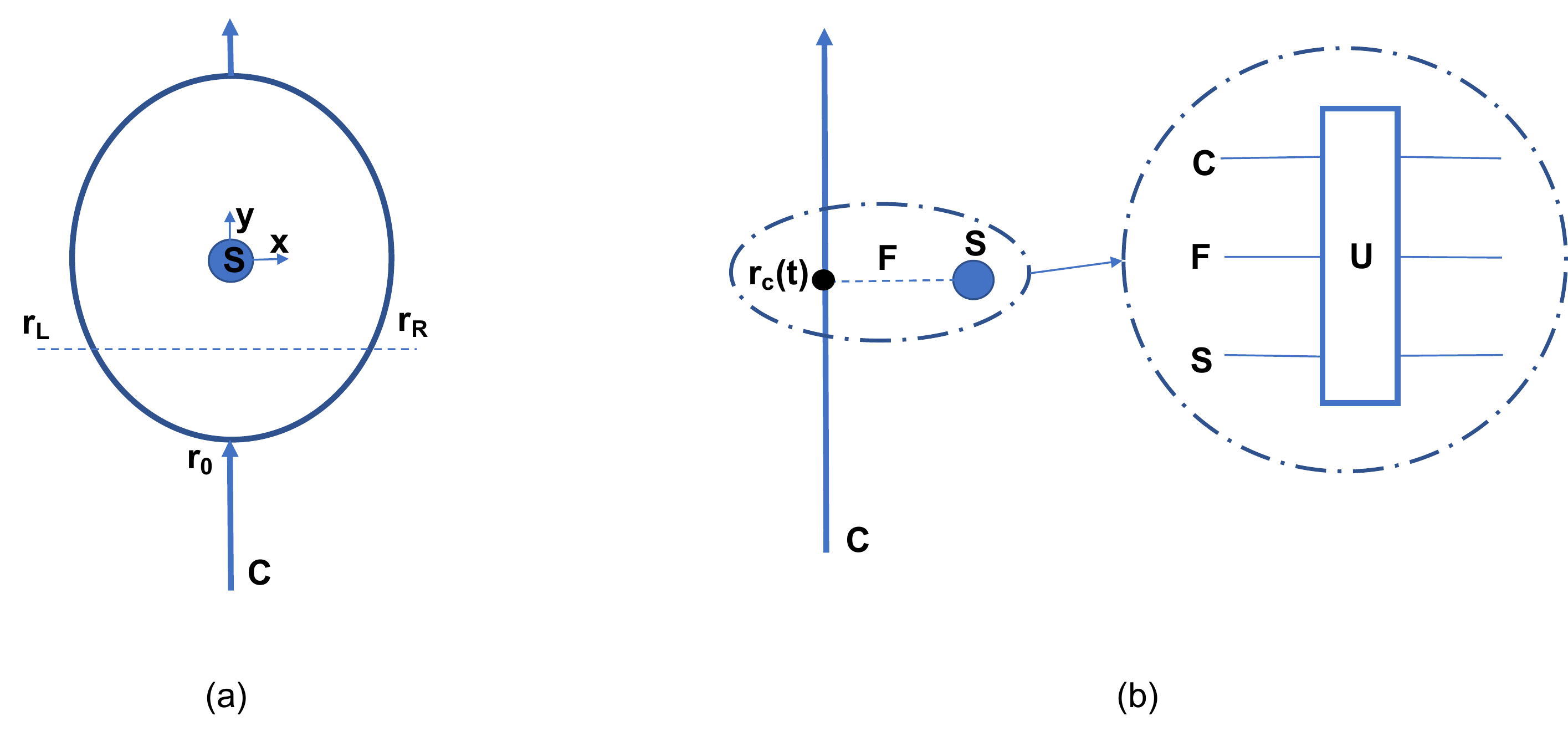}
\caption{(a): A Mach-Zehnder-like setup for the AB effect. C is the charge, S the solenoid. 
(b): Quantum network for the AB phase generation along the charge's path, via a local gate $U=\exp(iH_{AB})$ defined at each point ${\bf r_c(t)}$, involving the photon field.}
\label{PiC}
\end{figure} 

Now, the proposed field-based models are still non-local, because they do not explain how the phase, generated by interactions at the solenoid, is detectable by measuring observables of charge only. Therefore they solve problem (i), but not problem (ii). Here we address problem (ii) with a fully quantum model (where the EM field, the charge and the solenoid are quantised). Expanding on Vaidman's entanglement-based insight, we explain how the AB phase is mediated by the entanglement between the {\sl  photons} and the charge, achieved by local quantum coupling between the sources and the quantised EM field (which Vaidman's model does not describe, given that it treats the field as classical). We predict how the charge observables depend on the phase difference {\sl point by point} along the paths, bearing out Vaidman's \cite{VAI} and Kang's analyses \cite{KAN, KIM}, but via a fully local account, where the EM field is quantised.  Our model also vindicates (in quantum field theory) the conjecture that the locality issues in the AB effect are resolved by considering {\sl joint} gauge-transformations of both the charge field and the vector potential, \cite{WAL}. 

The key to addressing problem (ii) is to quantise the EM field. The interaction between the superposed charge and the quantised field fully accounts for how the AB phase is acquired locally, i.e., {\sl point-by-point along the charge's path}, just like for any other quantum phase. This local account is equivalent (not dramatically different \cite{LAT2}) to that of all EM phases.  Remarkably, our model produces a gauge-independent prediction for the phase difference at each point along the charge paths. Our prediction is testable: we propose an experiment to measure the phase difference, without closing the interference loop {\sl coherently}. We also address major problems of previously proposed schemes, \cite{KAN, KIM}, related to charge conservation or fermionic superselection rules \cite{KAN, KIM}. We obviate this problem by a state-tomography in the subspace of a two-charge system, compatible with the superselection rules.

{\bf The quantum model.}  The AB phase is generated by the quantised version of the classical problem where two sources interact electromagnetically, one of which (the charge in our case) is slowly varying, \cite{TAN}. 
Here we approximate the charge-solenoid interaction with two processes (adiabatic approximation): one is the charge's motion, with velocity $v<<c$, along a (possibly superposed) path; the other, defined for each point ${\bf r_c}$ along the charge path, is the process by which photons mediate the interaction between the charge and the solenoid, on the scale of the light speed $c$. We will focus on the latter, as it is the only relevant one. We model it by a phase gate $U$ which establishes the phase between two static sources (the charge at ${\bf r_c}$  and the solenoid at ${\bf r_s}$). Overall the effect {\sl not} static, because the charge distribution is non-stationary (albeit slowly varying).  

Consider the charge located at ${\bf r_c}$, the solenoid at ${\bf r_s}$ and the EM field $F$, whose observables act on the space ${\cal H}={\cal H}_C\otimes{\cal F}_R\otimes {\cal H}_C$, where each ${\cal H}_C$ is the Hilbert space of a single qubit and ${\cal F}$ denotes the Fock space of the photon field. 

We model the charge as a qubit, whose observables are generated by $(q_x^{(C)}, q_y^{(C)}, q_z^{(C)})$; $q_z^{(C)}$ represents the observable `whether the charge is on the left or on the right of the solenoid': its eigenstate $\ket{0}$ represents a sharp position at ${\bf r_L}$  (with eigenvalue $-1$) and $\ket{1}$ at ${\bf r_R}$ (with eigenvalue $+1$). The charge's dynamical evolution allows it to assume sharp values of $q_z^{(C)}$, but also sharp values of the complementary observable $q_x^{(C)}$ (when the charge is superposed across two points ${\bf r_L}$ on the left and ${\bf r_R}$ on the right of the solenoid): thus the charge is quantum. The solenoid is also modelled as a qubit, whose z component $q_z^{(S)}$ represents its presence/absence from the relevant point in the interferometer; but it is in the classical regime. Each component of the charge qubit is a generator of the Pauli algebra on ${\cal H}$, represented as $q_{\alpha}^{(C)}=\sigma_{\alpha}\otimes \mathbb{I}\otimes\mathbb{I}$ where $\sigma_{\alpha}$, $\alpha\in\{x,y,z\}$ is the element of the Pauli matrices operating on ${\cal H}_Q$; likewise, $q_{\alpha}^{(S)}=\mathbb{I}\otimes\mathbb{I}\otimes\sigma_{\alpha}$.

The field consists of N harmonic oscillators in momentum space, each mode $k$ represented by bosonic creation and annihilation operators $a_k, a_k^{\dagger}$, with $a_k=\mathbb{I}\otimes \hat a_k\otimes\mathbb{I}$ and $\hat a_k$ is the annihilation operator acting on the mode $k$ in $\cal{F}$ only.


The Hamiltonian (in the Coulomb gauge), \cite{TAN}, reads:

\begin{eqnarray}
H_{AB} & = &  E_C q_z^{(C)}  +  E_S q_z^{(S)}  \\
&+&  \int {\rm d^3{\bf k}} \hbar \omega_k a^\dagger_{k}a_{k} \label{AB} \\
& + &   \int {\rm d^3{\bf k}} g_k \frac{q}{m} {\bf p} \cdot {\bf u}_k (a_{k}e^{i {\bf k} {\bf r_c}}+a^\dagger_{k}e^{-i {\bf k} {\bf r_c}}) q_z^{(C)} \nonumber \\
& + &   \int {\rm d^3{\bf k}} \int {\rm d^3 {\bf x}}  g_k {\bf j}\cdot {\bf u}_k  (a_{k}e^{i {\bf k} {\bf x}}+a^\dagger_{k}e^{-i {\bf k} {\bf x}})q_z^{(S)}\nonumber \\
\end{eqnarray}

where $E_C$ and $E_S$ are the charge's and solenoid's free energies; $g_k=\sqrt{\frac{\hbar}{2\epsilon_0V\omega_k}}$ (V is the standard quantisation volume); $\omega_k$ and ${\bf k}$ represent the photon frequency and wavenumber of the $k$-th mode; ${\bf u}_k$ is the unit polarisation vector for mode k; ${\bf p}$ is the electron's momentum and ${\bf j}({\bf x}-{\bf r_s})$ is the solenoid's current distribution. One can recognise in the above formula the quantised vector potential: ${\bf A}({\bf x})=\int {\rm d^3{\bf k}} g_k {\bf u}_k (a_{k}e^{i {\bf k} {\bf x}}+a^\dagger_{k}e^{-i {\bf k} {\bf x}})\;.$ (We suppressed the polarisation index as it is irrelevant). 

Note that in the Coulomb gauge, the relation between the vector potential and the magnetic field is, at all times, {\sl non-local}: the vector potential at point $x$ is expressed as a function of the magnetic field at other points in space, see e.g. \cite{STEW}. Hence the reader might be concerned about our model being expressed in this non-local gauge. This is not a problem, because our aim is to show how a gauge-independent phase difference between any two points ${\bf r_L}$ and ${\bf r_R}$ is gradually acquired by the charge, in such a way that it is detectable by local action at each of these points. This is sufficient to prove that the AB phase is generated locally, like any other EM phase. We have chosen the Coulomb gauge for convenience; however, as we shall explain, the phase difference along the path has a gauge-independent expression: the same prediction could be made via any other gauge. Whether or not in each of these gauges the vector-potential (appearing in the Hamiltonian) has a local relation with the EM fields is not relevant for our discussion. The fact that it does not have a local relation to the magnetic field in the Coulomb gauge does not invalidate our claim: that the AB phase difference is locally and gradually acquired along the path, and detectable by acting locally on the charge.

{\bf The phase generation.}

The Hamiltonian acts on a fixed time interval $\tau$ (representing the time for light to travel from the charge to the solenoid and back again): $U=\exp{\left (-\frac{i}{\hbar}H_{AB}\tau\right)}$. Define $C_k=\frac{q}{m}g_k {\bf p} \cdot {\bf u}_k$ and $G_k=g_k {\bf j}({\bf x}-{\bf r_s}) \cdot {\bf u}_k$. Assuming that the charge is in a sharp position state $\ket{1}$ at ${\bf r_c}$ and the solenoid is in a sharp position state $\ket{1}$ at ${\bf r_s}$, the vacuum-to-vacuum transition amplitude is: 

\begin{eqnarray}
\langle1|_c \langle0|_F \langle1|_s\exp{(-i\frac{H_{AB}}{\hbar} \tau)}|1\rangle_c |0\rangle_F |1\rangle_s =\\ \nonumber
\exp\{-i\left(\xi+\phi({\bf r_c},{\bf r_s})\right)\tau\} \\ \nonumber
\end{eqnarray}

where: 

$$\xi=\frac{1}{\hbar}\int {\rm d}^3{\bf k} \int {\rm d}^3{\bf x} \left (E_C+E_S+4\frac{C_k^2+G_k^2}{\hbar\omega_k}\right)$$

is a phase which does not depend on the mutual position of solenoid and charge; while the position-dependent phase is:

$$\phi({\bf r_c},{\bf r_s})\doteq\frac{1}{\hbar}\int {\rm d}^3{\bf x}\int {\rm d}^3{\bf k}\left(8\frac{C_kG_k}{\hbar\omega_k} \cos{({\bf k}\cdot(\bf{r_c}-\bf{x}))}\right)\;.$$ 

As customary in quantum gates, we will set $\tau=1$ from now on. By noticing the identity:

$$
\int {\rm d}^3{\bf k}\left(8\frac{C_kG_k}{\hbar\omega_k} \cos{({\bf k}(\bf{r_c}-\bf{x}))}\right)=\frac{q}{m\epsilon_0c^2}\frac{{\bf p}\cdot{\bf j({\bf x}-{\bf r_s})}}{|\bf{r_c}-\bf{x}|}\;,
$$

we obtain the useful expression for the point-by-point phase of the charge: 
\begin{equation}
{\displaystyle{ \phi({\bf r_c},{\bf r_s})=\frac{1}{\hbar}{\cal E}({\bf r_c}, {\bf r_s})}}\;,\label{phase}
\end{equation}
where we have introduced the interaction energy between a charged particle and an infinite solenoid (a gauge-independent quantity): $${\cal E}({\bf r_c}, {\bf r_s})=\frac{1}{2}\int_V(\frac{{\bf B_0}{\bf B_c}}{\mu_0}+\epsilon_0{\bf E_s}{\bf E_c}){\rm d}^3r\;.$$ Here, ${\bf B_c}$ and ${\bf E_c}$ are the classical magnetic and electric fields generated by the charge located in ${\bf r_c}$; ${\bf B_0}$ and ${\bf E_s}$ are the electric and magnetic fields generated by an infinite solenoid positioned in ${\bf r_s}$.

In the approximation where the charge velocity $v$ is much lower than the light speed (\cite{BOY}):

$${\cal E}({\bf r_c}, {\bf r_s}) = \frac{qvB_0Sx}{2\pi(x^2+y^2)}$$

where $S$ is the solenoid cross-section; $x$ and $y$ are the cartesian coordinates of a coordinate system whose z-axis coincides with the normal to the solenoid cross-section and the y axis is parallel to the direction of motion of the electron before and after the interferometer, as represented in figure 1(a). 

Suppose the Hamiltonian acts on an initially superposed state between locations ${\bf r_L}$ and ${\bf r_R}$, where the location of the charge ${\bf r_c}$ is not sharp: $\ket{+}_c\ket{0}_F\ket{0}_s$, where $|+\rangle_c\doteq \frac{1}{\sqrt{2}}( \ket{0}_c+\ket{1}_c)$.  By linearity, $\displaystyle{\langle1|_c \langle0|_F \langle1|_s \left ( U\frac{1}{\sqrt{2}}\ket{+}_c\ket{0}_F\ket{0}_s\right)}$ depends on the phase difference
 
 \begin{equation}
\Delta\phi({\bf r_L}, {\bf r_R})\doteq \phi({\bf r_R},{\bf r_s})-\phi({\bf r_L},{\bf r_s})=\frac{2}{\hbar}|{\cal E}|\;.
\end{equation}

This is the phase difference available on the charge when it is superposed across any two points ${\bf r_R}$ and ${\bf r_L}$, as promised. It is a {\sl gauge-independent} quantity (\cite{KAN}), corresponding to the field energy variation due to the charge, point-by-point along the charge's path. 
The full AB phase $\Delta \Phi$, concurring with the standard approach, is obtained by integrating $\Delta \phi ({\bf r_L}, {\bf r_R})$ along a circular path, assuming $v=\frac{\pi\rho} {t_{loop}}$, where $\rho=\sqrt{(x^2+y^2)}$ is the radius of the circle and $t_{loop}$ is the total time taken by the charge to travel on the semicircle.
  
{\bf The Heisenberg picture.} In our model one can track how the x-component of the charge qubit directly depends on the phase, using the Heisenberg picture. Suppose that the initial values of the charge observables are $q_{\alpha}^{(C)}$, represented in terms of Pauli matrices. The Hamiltonian leaves $q_{z}^{(C)}$ unchanged, while the component $q_{x}^{(C)}$ changes as follows:
$$
q_{x}^{(C)} \rightarrow U^{\dagger}q_{x}^{(C)} U
$$ 

where $U=\exp{(-i\frac{H_{AB}}{\hbar} \tau)}$. Setting $\tau =1$,

\begin{eqnarray}
U^{\dagger}q_{x}^{(C)}U &=&\left(\alpha\cos{\theta} -\beta\sin{\theta} \right)  q_{x}^{(C)} \nonumber \\ 
&+& \left (\beta\cos{\theta} +\alpha \sin{\theta}\right )q_{y}^{(C)} \;,\\\nonumber 
\end{eqnarray}
where $$\displaystyle{\theta\doteq E_C+\phi({\bf r_c},{\bf r_s})q_{z}^{(S)}}\;,$$ $$\displaystyle{\alpha\doteq \cos \left(\int {\rm d}^3k \frac{C_k}{\hbar\omega_k}\left(\exp(-i{\bf k}{\bf r_c})a_k^{\dagger}+\exp(i{\bf k}{\bf r_c})a_k\right)\right)}$$ and $$\beta\doteq \sin\left(\int {\rm d}^3k \frac{C_k}{\hbar\omega_k}\left (\exp(-i{\bf k}{\bf r_c})a_k^{\dagger}+\exp(i{\bf k}{\bf r_c})a_k)\right )\right)\;.$$

By assuming the Heisenberg state to be $|+\rangle_c |0\rangle_F |1\rangle_s$, the expected value of the observable $U^{\dagger}q_{x}^{(C)}U$ is non-vanishing and depends on the phase difference $\Delta\phi({\bf r_L},{\bf r_R})$: thus measuring a function of this observable provides access to the phase along the path, without closing the loop coherently. 

{\bf An experimental proposal.} We now explain how to access the phase along the path by performing quantum tomography, with another reference charge. One has to define a procedure to extract the phase difference $\Delta \phi({\bf r_L},{\bf r_R})$ {\sl without closing the interferometer coherently}, because the latter would be tantamount to measuring the full AB phase. We adopt the picture of mode entanglement, (\cite{DUVE}), where the charge qubit consists of two spatial modes (left or right), each of which can contain 0 or 1 particles.  The Hadamard gate (figure 1) is an entangling operation between the path degree of freedom and the number of particles (0 or 1) on the path (\cite{DUVE}). 

Assume that the charge is an electron. Let $b_L^{\dagger}$, $b_L$  be fermionic creation/annihilation operators for it to be in a spatial mode $x_L$ on the left of S; and $b_R^{\dagger}$, $b_R$ be fermionic creation/annihilation operators for the electron to be in a spatial mode $x_R$ on the right.  The state where the charge is superposed across two locations is $\frac{1}{\sqrt{2}}(\ket{01}+\ket{10})$ where $\ket{01}=b_L^{\dagger}\ket{0}$, $\ket{10}=b_R^{\dagger}\ket{0}$ and $\ket{0}$ is the fermionic vacuum. Our hamiltonian will produce the state  $\frac{1}{\sqrt{2}}(\ket{01}+\exp(i\Delta\phi)\ket{10})$ where the phase $\Delta \phi$ is a function of the points ${\bf r_L},{\bf r_R}$ across which the electron is superposed and it is locally generated, as computed in our model.

Measuring the phase directly by local tomography on the charge only is impossible, because of fermionic and charge superselection rules, which impede measurements of observables such as $b_L+b_L^{\dagger}$, \cite{VAI2}. Crucially, the phase difference can still be reconstructed by utilising another reference electron, \cite{ENT}, and local tomography (on the left and right sides) involving the same number of electrons, without violating any superselection rule. This is an effective way of measuring the aforementioned x-component of the electron qubit, {\sl without closing the interferometer loop coherently} - i.e., with only decoherent communication between the two sides, thus guaranteeing that the measured phase is not a closed-loop type phase.

Consider a reference electron (labelled as B) superposed across the two paths, which does {\sl not} pick up the AB phase (unlike electron A, which does), as follows. 

\begin{figure}[htb]
\centering
\includegraphics[width=85mm]{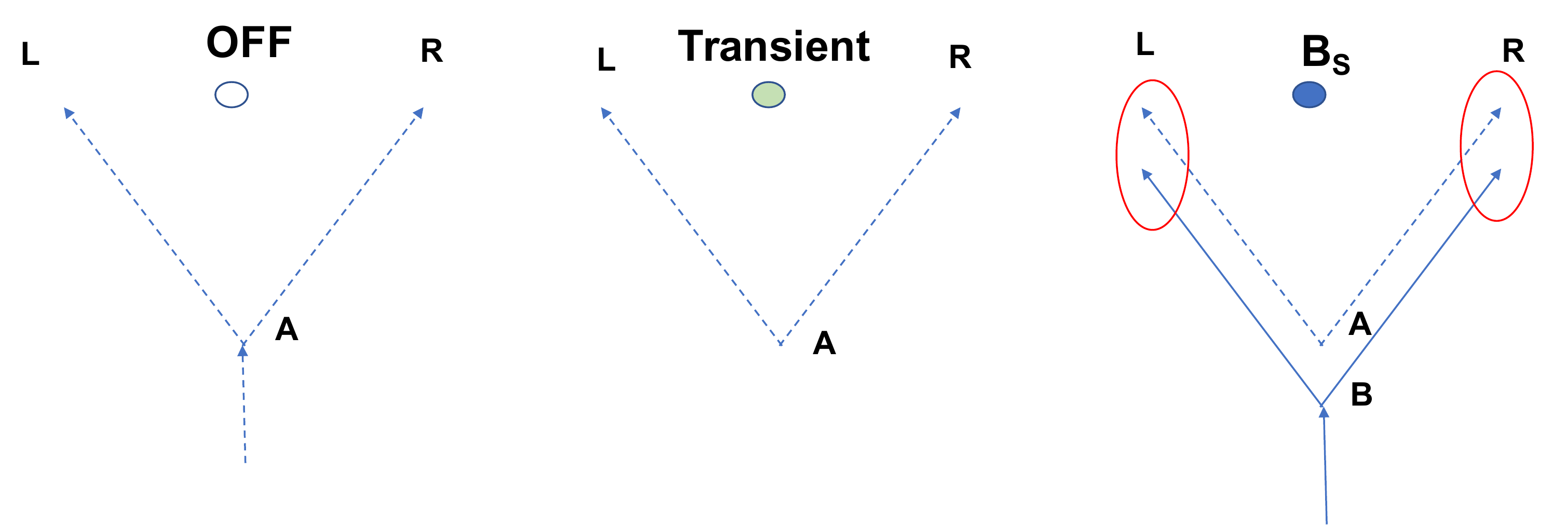}
\caption{{\bf Left}: The reference electron A is superposed across the left (L) and right (R) modes; the solenoid is off. {\bf Center}: the solenoid gradually reaches the desired current value; electron A acquires a fixed relative phase. 
{\bf Right}: Electron B is superposed across L and R, acquiring the AB phase. Joint measurements of A and B, local to the left and right modes, reconstruct the partial AB phase.}
\label{PiC}
\end{figure}

Suppose electron B and the electron A are both in the lower semi-plane defined by a line passing through the solenoid. The solenoid is initially switched off and electron B is superposed across two locations. Then the solenoid is switched on and brought to the desired stationary current to cause the AB effect on A. This transient will cause a relative phase to appear on B. But 1) this phase is not AB-like (because it is induced locally on B by the non-zero electric field produced by the transient current); 2) it is a fixed controllable offset, so we can in principle set it to zero.  Electrons A and B will be therefore in the state: 

$$
\frac{1}{{2}}(\ket{0_L1_R}_A+\exp(i\Delta \phi)\ket{1_L0_R}_A)(\ket{0_L1_R}_B+\ket{1_R0_L}_B)
$$
where $\exp(i\Delta \phi)$ is the AB phase difference across the two points $({\bf r_L},{\bf r_R})$ along the path of the electron. One can group the terms relative to the left and to the right modes, as follows:

\begin{eqnarray}
&\frac{1}{{2}}\times&\ket{0_A0_B}_L\ket{1_A1_B}_R+\exp(i\Delta \phi)\ket{1_A1_B}_L\ket{0_A0_B}_R \nonumber \\
&+&\ket{0_A1_B}_L\ket{1_A0_B}_R+\exp(i\Delta \phi)\ket{1_A0_B}_L\ket{0_A1_B}_R\;.\nonumber \\
\end{eqnarray}

In the branches where only one charge is present on the left and right arms (second line of the above equation) the phase can be detected by measuring, locally on the left and on the right, observable whose eigenstates are superpositions of $\ket{0_A1_B}_L$ and $\ket{1_A0_B}_L$; likewise for the right side. Measuring these observables does not violate charge conservation or fermionic superselection rules. By local tomographic reconstruction of the 1-particle sector of the above state, one can retrieve the phase difference at any point along the path, without closing the interferometer coherently, as promised.  This protocol differs from that in \cite{REZ}, because the latter measures the full phase AB coherently, by requiring that the electron and a positron imprint it onto a photon by annihilating half-way through the interference experiment. Here, instead, the electrons A and B do not enclose the solenoid coherently. A similar conclusion could be reached with other experiments. For instance, if light took longer to complete a round trip between the electron and the solenoid, compared to time for the electron to perform full interference, our model predicts that no AB phase would be observed, while the semiclassical models would predict that the phase should be observed. This could be tested in principle by inserting a material that slows photons appropriately between the electron and the solenoid.   

{\bf Discussion.} Observing the locally built up phase refutes the idea that the AB phase is anomalous because of its non-locality, \cite{LAT2}.  An experiment implementing the suggested tomographic reconstruction of the partial phase would rule out all models maintaining that the phase is created non-locally and that it is observable only once the path is closed. It could not be explained by the semiclassical field-based models (\cite{VAI, KAN, WAL}) either, because they do not explain how the phase, generated at the solenoid, travels to the electron where it is detected. 

The local phase built-up does not contradict the fact that the EM observables must be functions of (gauge invariant) quantities, i.e. integrals of the vector potential along closed loops, such as $e^{iq\oint Adx}$. For any fraction of this phase is also observable, e.g. its $n$-th root, $\sqrt[n]{\exp\{iq\oint {\bf A}{\bf dl}\}}$. This fractional phase is acquired during the journey of the charge on the path that is $n$ times smaller than the one that is required to close the interferometer (assuming that the charge travels at a constant speed). 

Our model shows that the AB effect is important not because it is based on a non-local mechanism, but because it is the unique case where models treating the EM field as a classical background are inadequate - they all require some non-locality. Given the role of entanglement in the AB phase, \cite{VAI}, modelling the field as a classical background is bound to lead to apparent non-locality. As recently proved, a classical mediator cannot locally induce entanglement between two systems, \cite{MAVE}. Also, if the EM field is modelled as a collection of quantum harmonic oscillators, one can no longer say that the field ``is {\sl zero}" at a particular point. Even when the EM field is in its vacuum state (the expected values of the field components are zero), its observables consist of (non-zero) q-numbers, locally coupled with the charge's and the solenoid's q-numbered observables. We conjecture that our quantum treatment can explain {\sl some} of the variants of the AB effect experiment, \cite{LAT1}. This is because it explicitly includes the interaction of the charge with photons. 

As argued, there are two different problems: (i) whether there is a model for the AB phase expressed in terms of fields only; (ii) whether the AB phase is generated {\sl locally as all other EM phases}, i.e. whether it is built up gradually along the charge's path. Vaidman's model addresses (i), but still has non-locality. Our model addresses (ii), using a model with the (quantum) vector potential, in the adiabatic approximation. An outstanding problem is finding a {\sl local} quantum-field theory hamiltonian expressed with fields only. However, this is not a special issue arising in the AB effect: it affects all EM hamiltonians with interactions. 

Our experimentally testable quantum model has thus lifted the doubt on whether the AB phase is generated by a special, non-local type of EM interaction. As we explained, it is not.

\textit{Acknowledgments}: The Authors thank D. Deutsch, A. Ekert and L. Vaidman for discussions; H. Brown, K. Burnett, E. Cohen, K. Kang, S. Kuypers, P. Saldanha, B. Yadin and A. Tibau Vidal for several helpful comments. CM thanks the Templeton World Charity Foundation and the Eutopia Foundation. This research is supported by the National Research Foundation, Prime Minister's Office, Singapore, under its Competitive Research Programme (CRP Award No. NRF- CRP14-2014-02) and administered by Centre for Quantum Technologies, National University of Singapore.

\end{document}